\documentstyle[twocolumn,epsf]{jpsj}

\def\runtitle { Phase Changes in an Inelastic Hard Disk System with a Heat Bath under Weak Gravity for Granular Fluidization }

\def\runauthor{Masaharu {\sc Isobe} and Hiizu {\sc Nakanishi}}

\title { Phase Changes in an Inelastic Hard Disk System with a Heat Bath under Weak Gravity for Granular Fluidization }

\author
{ 
Masaharu {\sc Isobe}\footnote{E-mail: isobe@stat.phys.kyushu-u.ac.jp.}
 and Hiizu {\sc Nakanishi}
}

\inst
{
Department of Physics, Kyushu University 33, Fukuoka 812-8581
}

\recdate
{
April 16, 1999
}

\abst{
We performed numerical simulations on a two-dimensional inelastic hard disk system under gravity with a heat bath to study the dynamics of granular fluidization.
Upon increasing the temperature of the heat bath, we found that three phases, namely, the {\it condensed phase}, {\it locally fluidized phase}, and {\it granular turbulent phase}, can be distinguished using the maximum packing fraction and the excitation ratio, or the ratio of the kinetic energy to the potential energy.
It is shown that the system behavior in each phase is very different from that of an ordinary vibrating bed.
}

\kword{ inelastic hard disk, event-driven molecular dynamics simulation, non-equilibrium steady state, granular fluidization, velocity PDF, flatness parameter}

\begin{document}
\sloppy
\maketitle

The dynamics of fluidized granular systems has attracted much attention in physics communities as a nonequilibrium statistical system\cite{hayakawa95,jaeger96,kadanoff99}.
Due to the fact that an element particle in granular systems is already macroscopic, there are two major differences in their dynamics from that of an ordinary molecular system.
First, thermal fluctuation does not play any role because relevant energy scales for the kinetic and potential energy are much larger than the thermal energy.
Second, the dynamics is dissipative because the degrees of freedom we are dealing with are coupled with microscopic processes that are not treated explicitly.
Because of these properties, granular systems require an energy source in order to be in a steady state.
It is also expected that the external gravitational force plays an important role in their dynamics.
In most experimental situations where the systems are excited by a vibrating plate, the effect of gravity is very large.\cite{knight93,clement92,evesque89,taguchi92a,gallas92,aoki96}
It has been found that various patterns and localized oscillations appear as the intensity and frequency of the vibration are changed.\cite{MUS95,UMS98,UMS96}
It has been also demonstrated that the velocity distribution of the particles does not have the Maxwell-Boltzmann distribution, reflecting the fact that the system is not in equilibrium.\cite{taguchi95,ichiki95,murayama98}
Systems devoid of gravity are also being examined mainly by computer simulations and theoretical analyses.\cite{goldhirsch93,mcnamara96,deltour97,grossman97,peng98,PLMV-cond98,du95}
A one-dimensional system with a heat bath at one end has been investigated and it has been found that the spatial distribution of particles is singular, therefore, the hydrodynamical description of the behavior does not appear to be possible.\cite{du95}
Systems with uniform excitation, in which all the particles are agitated by the Langevin force, are also studied\cite{peng98,PLMV-cond98} because these system are statistically homogeneous and have a thermodynamic limit, even though such a situation may not be easily realized experimentally.
Large density fluctuation and non-Gaussian velocity distribution have been observed in such systems\cite{PLMV-cond98} when the effect of particle collision is larger than that of the Langevin force.
Since the gravitational force should have a large effect on clustering, it would be of interest to observe how the system behavior changes as the gravity sets in, which is one of the motivating factors behind this work.
An experimental study for analyzing the relative effects of gravity to those of excitation has been reported.\cite{kudrolli97}
In the experiment, stainless steel balls were placed on a plate that was held almost horizontally, and were excited by the vibrating wall of the lower side.
By tilting the system, the effective gravity was changed and it was shown that even a small force of gravity makes the cluster migrate downward.
In this letter, we examine the effect of gravity on granular fluidization by numerical simulations.
It is demonstrated that a system undergoes a few phase changes as the ratio of the external driving to the gravity is changed.

The system considered here consists of hard disks of mass $m$ and diameter $d$ in a two-dimensional space under uniform gravity.
We employ the periodic boundary condition in the horizontal direction and the system is considered to be high enough to prevent the particles from hitting the ceiling too frequently.
The particles collide with each other with a restitution constant $r$.
All the disks are identical, namely the system is monodispersed.
For simplicity, we neglect the rotational degree of freedom.
Because of the hard core interaction, collision is instantaneous and only binary collisions occur.
When two disks, $i$ and $j$, with respective velocities ${\bf u}_i$ and ${\bf u}_j$ collide, the velocities after the collision, ${\bf u}_i'$ and ${\bf u}_j'$, are given by
\begin{eqnarray}
{\bf u}_i' & = &
{\bf u}_i-\frac{1}{2}(1+r)[{\bf n}\cdot ({\bf u}_i-{\bf u}_j)]{\bf n}
\\
{\bf u}_j' & = &
{\bf u}_j+\frac{1}{2}(1+r)[{\bf n}\cdot ({\bf u}_i-{\bf u}_j)]{\bf n},
\end{eqnarray}
where ${\bf n}$ is the unit vector parallel to the relative position of the two colliding particles in contact.
Between the colliding events, the particles undergo free fall motion with the gravitational acceleration ${\bf g}\equiv (0,-g)$ following parabolic trajectories.
The system is driven by a heat bath with temperature $T_w$ at the bottom of the system; a disk hitting the bottom bounces back with velocity ${\bf v}=(v_x,v_y)$ chosen randomly by the probability distributions $\phi_x(v_x)$ and $\phi_y(v_y)$;
\begin{eqnarray}
\phi_x(v_x) & = &
\sqrt{\frac{m}{2\pi k_B T_w}} e^{-mv_x^2/2k_B T_w}
~ (-\infty < v_x < \infty)
\label{eq:heat1}
\\
\phi_y(v_y) & = &
\frac{m}{k_B T_w} v_y e^{-m v_y^2/2k_B T_w}
~\quad (0<v_y<\infty) ,
\label{eq:heat2}
\end{eqnarray}
where $k_B$ is the Boltzmann constant\cite{tehver98}.
We assume that the collision with the ceiling is elastic with an unchanged horizontal velocity.
Note that the wall temperature $T_w$ here is just a parameter to characterize the external driving and is not related to the thermal fluctuation.
We call this system ``inelastic hard disk system with a heat bath under weak gravity'' (IHSHG) to emphasize the importance of the competition between excitation of the  heat bath and gravity.
The system is completely characterized with only four dimensionless parameters; the restitution coefficient $r$, the driving intensity $\Lambda \equiv k_BT_w/mgd$, the system width in the unit of disk diameter $N_w\equiv L/d$, and the number of layers $N_h\equiv N/N_w$, where $L$ is the system width and $N$ is the total number of disks.
The present system is analogous to the ordinary granular vibrated bed\cite{gallas92,taguchi92a,aoki96,taguchi95,murayama98}, but simpler because it does not have an external time scale, or a period of external driving vibration.
We employ the event-driven molecular dynamics method to simulate the IHSHG model; we have developed a simple and efficient algorithm\cite{isobe99,rapaport80,marin93}, and achieved the fastest simulation speed in the world (about 460 million collisions per CPU hour for the 4000 disk system on a VT-Alpha-600), which allows us to simulate the system over a wide range of parameters.
Most of the simulations presented in this letter were performed on the system with $r=0.9$, $N_w=200$, and $N_h=20$ (i.e. $N=4000$).
The driving intensity $\Lambda$ was varied from 110 to 770 to study the change in the system behavior.
To ensure that the system had reached a steady state, we relaxed the system until the total energy did not drift.
Thereafter, we started the simulations with the length of 10,000 collisions per particle to obtain the data.
Inelastic collapse\cite{mcnamara94} did not occur in the present system during the simulation time for the parameter region studied here.

In Fig.~1, a typical snapshot of the simulation is shown for $\Lambda=182$ with $(r,N_w,N_h)=(0.9,100,20)$.
The average packing fraction profile as a function of height $y$ is also shown.
It can be seen that the density is small near the bottom because of the excitation by the heat bath, and the packing fraction reaches the maximum value $A_{\rm max}$ at a certain height.
\begin{figure}
\begin{center}
\leavevmode
\epsfxsize=8.0cm
\epsfbox{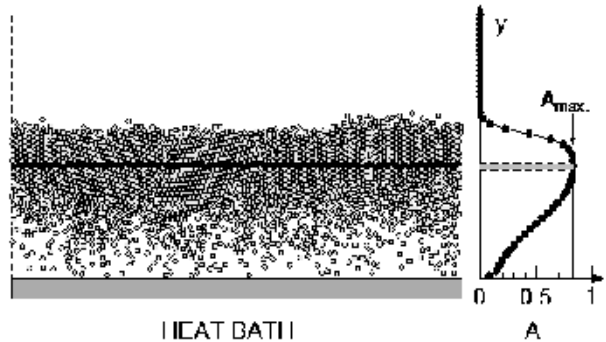}
\end{center}
\caption{
A typical snapshot of the simulation and the average packing fraction profile as a function of height $y$.
The parameters are $(r,\Lambda, N_w, N_h)=(0.9,182,100,20)$.
The black disks are the disks in the layer where the packing fraction reaches a maximum value.
}
\label{fig.1}
\end{figure}

In order to characterize the steady states, we measure the maximum packing fraction $A_{\rm max}$ as a function of the driving intensity $\Lambda$ ($\circ$ in Fig.~2).
There are two cusps around $\Lambda=200$ and 380, suggesting phase transitions with changing $\Lambda$.
These transitions should be related to the excitation structure of the state, therefore we define the excitation ratio $\mu$ by $\mu\equiv{\cal K}/{\cal U}$, namely, the ratio of the total kinetic energy ${\cal K}=m/2\sum_i v_i^2$ to the potential energy ${\cal U}=mg\sum_i y_i$.
The cusps appear at the same points for $\mu$, which confirms the transitions ($\times$ in Fig.~2).
\begin{figure}
\begin{center}
\leavevmode
\epsfxsize=7.0cm
\epsfbox{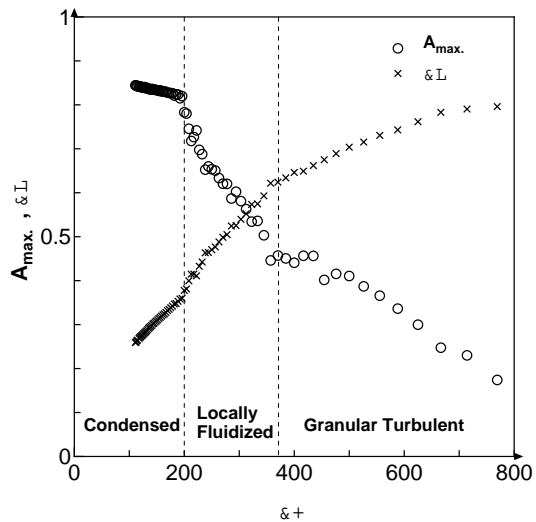}
\end{center}
\caption{
The maximum packing fraction $A_{max}$ and the excitation ratio $\mu$ vs. the driving intensity $\Lambda$.
The system parameters are $(r, N_w, N_h)=(0.9,200,20)$.
The regions of $\Lambda$ for the condensed phase, the locally fluidized phase, and the granular turbulent phase are shown.
}
\label{fig.2}
\end{figure}

In order to conceptualize the underlying physical mechanism of the system behavior in each phase separated by the above transitions, three snapshots for typical values of $\Lambda$ for each phase are shown in Fig.~3.
\begin{figure}
\begin{center}
\leavevmode
\epsfxsize=8.0cm
\epsfbox{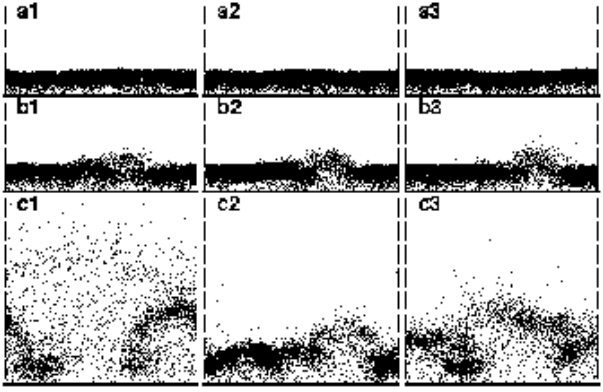}
\end{center}
\caption{
The series of snapshots for the three typical values of $\Lambda$, a1--a2 ($\Lambda=182$), b1--b2 ($\Lambda=250$), c1--c2 ($\Lambda=667$).
The system parameters are $(r,N_w,N_h)=(0.9,200,20)$.
}
\label{fig.3}
\end{figure}
For $\Lambda=182$ (a1--a3 in Fig.~3), there is a closed packing layer and the state is not very dynamic, because excitation is low and the potential energy is dominant.
We call the phase in $\Lambda\leq 200$, the {\it condensed phase}.
It is apparent, however, that collective motion appears on the surface of the layer.
In the second phase, $\Lambda=250$ (b1--b3 in Fig.~3), the closed packing layer is locally broken by excitation of the heat bath.
The high-speed particles are blown upward from the holes in the layer.
In Fig.~3, we see only one hole in the system, but for a larger system, there are certain cases where we can observe more than one hole.
The holes migrate and occasionally become more active; sometimes they almost close temporarily, but the structure is fairly stable.
We call this phase the {\it locally fluidized phase}.
For the case of $\Lambda=667$ in the third phase (c1--c3 in Fig.~3), the layer is completely destroyed.
The average density is quite low, but it is very different from the ordinary molecular gas phase.
The density fluctuation is very large and this fluctuation causes turbulent motion driven by the gravity.
At some time, the whole system gets excited with some relatively smaller density fluctuations, but the very next moment, a large proportion of the particles travels downward and forms a layer like structure.
This structure, however, is destroyed immediately.
We call this phase the {\it granular turbulent phase}.

These inhomogeneous behaviors should result in a non-Maxwell-Boltzmann distribution of velocity.
Figure~4(a) shows the horizontal velocity distribution functions at the horizontal layer around the height of the maximum packing fraction for each phase in the log-linear scale.
The distributions for $\Lambda$=250 and 667 deviate from the Gaussian and are more or less in exponential form in the tail region.
It can be seen that the distribution for $\Lambda=250$ in the locally fluidized phase consists of two parts; the central part that originates from the closed packing layer, and the wide tail that comes from the fluidized holes.
The central part is very close to that for $\Lambda=182$.
\begin{figure}
\begin{center}
\leavevmode
\epsfxsize=7.0cm
\epsfbox{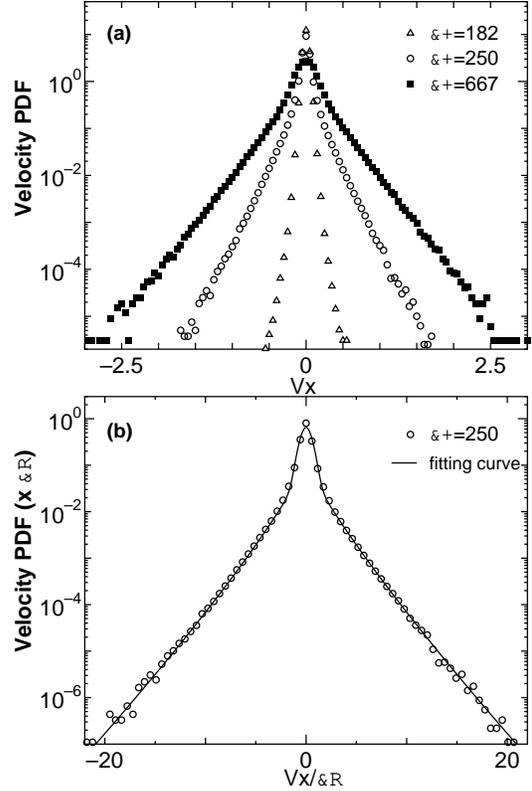}
\end{center}
\caption{
The horizontal velocity probability distribution functions
 for the three typical values of $\Lambda$.
The other parameters are $(r,N_w,N_h)=(0.9,200,20)$.
(a) The data are plotted in the unit of $\sqrt{k_BT/m}$.
(b) Equation~(\ref{phi}) is fitted to the data for $\Lambda=250$
 with the parameters $\sigma_G=0.048$, $x_0/\sigma_G=1.7$, $\beta=0.85$, and $p=0.25$.
Note that the data are scaled by the standard deviation $\sigma$.
}
\label{fig.4}
\end{figure}

In order to quantify the deviation, we calculated the flatness parameter $f\equiv <v_x^4>/<v_x^2>^2$ (Fig.~5).
Over the whole region, the value of $f$ is different from 3, which is the value for the Gaussian distribution, but it is remarkable that $f$ becomes very large, as large as 20, in the {\it locally fluidized phase}.
This unexpected large value of $f$, however, can be understood naturally as follows.
Assume the PDF $\phi(v)$ has two components, the narrow Gaussian distribution $\phi_{G}(v)$ with the weight~$1-p$ and the broader stretched exponential distribution $\phi_{S}(v)$ with the weight~$p$;
\begin{equation}
\phi(v) = p \phi_{S}(v) + (1-p) \phi_{G}(v),
\quad (0\le p\le 1),
\label{phi}
\end{equation}
where $\phi_S(x)\equiv (\beta /[2x_0\Gamma(1/\beta)])e^{-|x/x_0|^\beta}$ and
 $\phi_G(x)\equiv [1/(\sqrt{2\pi}\sigma_G)]e^{-x^2/2\sigma_G^2}$.
Then, the flatness of this distribution is given by
\begin{equation}
f = { [\Gamma(5/\beta)/\Gamma(1/\beta)]\tilde x_0^4 p +3(1-p)
\over \bigl(
[\Gamma(3/\beta)/\Gamma(1/\beta)]\tilde x_0^2 p+ (1-p)
\bigr)^2};
\quad
\tilde x_0 \equiv {x_0\over\sigma_G}
\label{flatness}
\end{equation}
because the second and the fourth moment of $\phi_S(x)$ are given by $\Gamma(3/\beta)/\Gamma(1/\beta)\cdot x_0^2$ and $\Gamma(5/\beta)/\Gamma(1/\beta)\cdot x_0^4$, respectively.
If we fit the parameters in eq.~(\ref{phi}) to the data for $\Lambda=250$, we obtain $\sigma_G=0.048, \tilde x_0=1.7$, $\beta=0.85$, and $p=0.25$, which yields $f\simeq 19.0$ (Fig.~4(b)).
The enhancement of the flatness by the superposition of the broader distribution can even be drastic for a small value of $p$ as can be seen if eq.~(\ref{flatness}) is plotted as a function of $p$.
This observation indicates that $f$ can be used as a sensitive index to detect the appearance of a small weight of the broad component in the distribution.
The sharp rise of $f$ in Fig.~5 around $\Lambda=200$ is clear evidence of the appearance of the fluidized holes.
\begin{figure}
\begin{center}
\leavevmode
\epsfxsize=7.0cm
\epsfbox{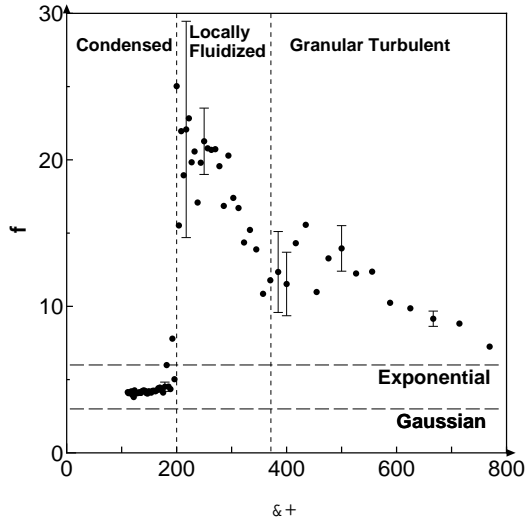}
\end{center}
\caption{
The flatness $f$ vs. the driving intensity $\Lambda$.
The other parameters are $(r,N_w,N_h)=(0.9,200,20)$.
The values for the Gaussian distribution $(f=3)$ and the exponential distribution $(f=6)$ are indicated in the figure.
The statistical error bars with twice the standard deviation are also plotted in the several typical data for each phase.
}
\label{fig.5}
\end{figure}

There are some interesting issues regarding the dynamics in each phase.

(1) What is the mechanism of the surface wave-like motion in the {\it condensed phase}?
There must be no surface tension in the granular system, therefore the restoring force should originate from the balance between the gravity and the excitation, but it is not clear if it could be described as an ordinary gravitational wave in fluid.

(2) The localized excitation in the {\it locally fluidized phase} should resemble a circle in a 3-d system and reminds us of an oscillon,\cite{UMS96} but they are different; the external vibration frequency is essential for the oscillon dynamics, but the localized excitation here does not have such a characteristic frequency.
In b1--b3 of Fig.~3, we can observe the merging process of two excitations, but their mode of interaction is not clear.
An interesting issue here is that if the transition from the {\it fluidized phase} to the {\it condensed phase} and/or the transition to the {\it granular turbulent phase} can be understood in terms of the local excitation, does the {\it locally fluidized phase} transform to the {\it condensed phase} when the distance between the excitations diverge?
Does the {\it locally fluidized phase} transform to the {\it granular turbulent phase} at the point where they condense?

(3) Turbulent motion does not exist in the 1-d hard rod system\cite{PLMV-cond98}.  It should be very different from ordinary fluid turbulence.
In the finite system with finite height, turbulence may disappear when the driving $\Lambda$ is large enough, but in the system with infinite height, it appears that turbulent motion persists however large the driving $\Lambda$ is.

Before concluding, let us discuss the relationship between the present model and the ordinary vibrated bed, where the system is driven by a vibrating plate at a given frequency.
In such a case, the control parameter is usually taken as the ratio of the accelerations $\Gamma\equiv A\omega^2/g$, where $A$ and $\omega$ are the amplitude and the angular frequency of the vibration, respectively.
In most experimental situations, the external frequency dominates the system time scale and the collision interval is directly related to $1/\omega$.
In the present case, the control parameter is the ratio of the energies $\Lambda =k_BT_w/mgd$ and no external time scale is imposed.
This situation may correspond to that in the vibrated bed when $1/\omega$ is much shorter than any other of the relevant time scales in the system.

In summary, we performed numerical simulations for a two-dimensional inelastic hard sphere system with a heat bath under gravity.
Upon increasing the heat bath temperature, the system exhibited three distinct phases, namely, the {\it condensed phase}, the {\it locally fluidized phase}, and the {\it granular turbulent phase}.
Their dynamics are very different from those of not only an ordinary molecular system but also of a conventional vibration bed system.

We would like to thank Professor Hayakawa for helpful discussions.  A part of the computation in this work was performed using the facilities at the Supercomputer Center, Institute for Solid State Physics, University of Tokyo.

\end{document}